\documentclass[preprint,floats,twoside,tightenlines,eqsecnum,aps,prd,
showpacs,showkeys,floatfix]{revtex4}

\usepackage{graphicx}
\usepackage{bm}

\newcommand{\beqa} {\begin{eqnarray}}
\newcommand{\eeqa} {\end{eqnarray}}
\newcommand{\nn} {\nonumber}
\newcommand{\rf}{\ref}
\newcommand{\ct}{\cite}
\newcommand{\lb}{\label}

\newcommand{\tr}{{\rm tr }}

\newcommand{\beq}{\begin{equation}}
\newcommand{\enq}{\end{equation}}
\newcommand{\beqast}{\begin{eqnarray*}}
\newcommand{\enqa}{\end{eqnarray}}
\newcommand{\enqast}{\end{eqnarray*}}

\newcommand{\cD}{{\cal D}}

\newcommand{\cS}{{\cal S}}

\newcommand{\bB}{\mbox{\bf B}}
\newcommand{\bF}{\mbox{\bf F}}

\newcommand{\al}{\alpha}

\newcommand{\ga}{\gamma}
\newcommand{\de}{\delta}
\newcommand{\ep}{\epsilon}

\newcommand{\ka}{\kappa}
\newcommand{\lam}{\lambda}
\newcommand{\rh}{\rho}
\newcommand{\si}{\sigma}

\newcommand{\om}{\omega}

\newcommand{\Ga}{\Gamma}

\newcommand{\lag}{\langle}
\newcommand{\rag}{\rangle}
\newcommand{\half}{{\textstyle \frac{1}{2}}}

\begin{document}
\title{ Nonperturbative QCD treatment of $J/\Psi$ photoproduction }
\author{H.G. Dosch }
\affiliation{Institut f\"ur Theoretische Physik, Universit\"at Heidelberg\\
 Philosophenweg 16, D-6900 Heidelberg, Germany }
\author{E. Ferreira}
\affiliation{Instituto de F\'{\i}sica, Universidade Federal do Rio de
Janeiro \\
C.P. 68528, Rio de Janeiro 21945-970, RJ, Brazil   }

\date{\today}

\begin{abstract}
\noindent
We present a nonperturbative QCD calculation of elastic $J/\psi$  meson
production in photon-proton scattering at high energies.
Using light cone wave functions of the photon and vector mesons, and
the framework of the  model of the stochastic QCD vacuum,  we
calculate the differential and integrated elastic cross sections
for  $\gamma~  p \rightarrow J/\psi~  p $. With an  energy
dependence
 following the two-pomeron model
we are able to give a consistent description of the integrated cross sections
and the differential cross sections at low $|t|$  in the range from 20
GeV up to the highest HERA energies. We discuss different approaches to
introduce saturation and find no specific effects up to energies presently
available. We also calculate and compare to experiments the cross section 
for $\Upsilon$ photoproduction. 
 \end{abstract}
 \bigskip


 \maketitle

\today

\section{Introduction}


Photoproduction of J/$\psi$ mesons is at the borderline of soft
and hard physics. On one side the mass of the charmed quark provides
a relatively large scale, but on the other hand the size of the
J/$\psi$ meson is determined not only by the mass of the charmed
quark but also by the confinement mechanism  which therefore cannot 
be neglected. Indeed if confinement effects could be totally neglected
the size of the $J/\psi$ would be of the order of the Coulomb radius
 $1/(m_c \alpha_s)$ which is 
considerably larger than the Compton-wave length $1/m_c$.

J/$\psi$ photoproduction has been treated  extensively with methods
of perturbative QCD~\cite{pqcd1,pqcd2,pqcd3,pqcd4,pqcd5}. In this paper we
present a nonperturbative approach.
This has the disadvantage of stronger model dependence, but the
advantage that the process can be viewed from
an unified point of view together with other processes already studied,
and no use of external quantities like parton distributions is needed.
The only intervening quantities are
inherently calculated in the nonperturbative approach and no new
free parameters have to be introduced.
By comparing the successes and limitations of the perturbative and
nonperturbative approaches, important insight in the transition
region between the two QCD regimes can  be obtained.

The approach presented here is based on a functional integral
treatment of high energy scattering \cite{Nac91}
where the functional integrals are evaluated in an extension of the
stochastic vacuum model \cite{Dos87,DS88}. The method  has been
applied with great success to calculate differential
and total cross sections for many processes. 

Although the size of the $\Upsilon$ is much smaller than that of
the $J/\psi$ and hard contributions are expected to be
dominant, we nevertheless also calculate  photoproduction
of the $\Upsilon$-meson in our model, obtaining reasonable
agreement with experiment.

Our paper is organized as follows. In Sect. 2 we discuss shortly
the main features of  the underlying nonperturbative model  
and present the  final formulae for dipole-dipole
scattering, which is the basic ingredient in our approach. We also
give convenient parametrisations of our theoretical results  and
discuss the  inherent limitations of the model. Our numerical
results and a comparison with experiment are given in Sect. 3.
The paper closes with the discussion in Sect. 4. In the Appendix
we collect some useful formulae concerning wave functions.
\section{The model}
\subsection{Nonperturbative treatment of scattering amplitudes}

In this subsection we give a short review of the main ideas
behind our  nonperturbative model for soft high energy reactions
and present the final formulae. For more details we refer to the
original literature \cite{Nac91,DFK94} and reviews
\cite{Nac96,Dos96,Dos99}. The functional integral  approach to
soft high energy scattering~\cite{Nac91}  starts from the scattering
of a highly energetic quark in an external colour field $\bB_\mu(x)$.
Along its path $\Gamma$ the quark  picks up the non-Abelian phase
\[e^{-ig \int_\Gamma \bB dx} ~ ,\] where the expression is path ordered.
Here and in the following we express through bold face letters matrix
valued quantities like
\beq
\bB_\mu(x) = \sum_{C=1}^8 \frac{1}{2}\lambda_C B^C_\mu(x) ~ ,
 \enq
where $\lambda_C$ represents the Gell-Mann matrices.

\begin{figure}[ht]
\vskip 2mm
\includegraphics[height=10cm,width=10cm]{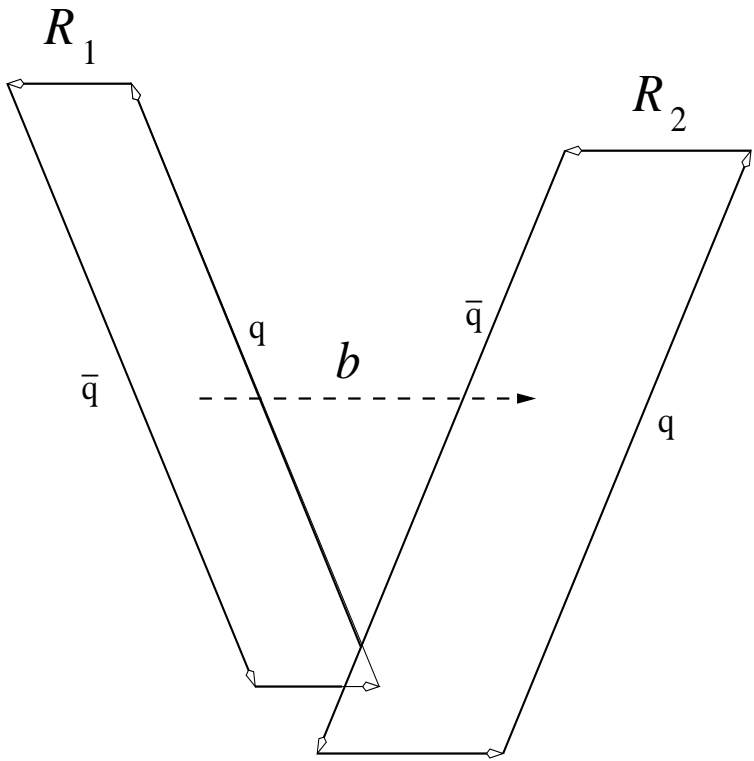}
\caption{\label{twoloops} The scattering of two dipoles.}
\end{figure}

 According to the functional integral approach to quantisation
 the scattering amplitude of two quarks can be obtained by
averaging these phase factors of  two quarks with the
 exponential of the action as weight.
 Formally this can written as the functional integral
 \beq
\int \cD B e^{-ig \int_{\Gamma_1}\bB dx}e^{-ig \int_{\Gamma_2}\bB dx}
\exp[-i S_{QCD}] \equiv
 \lag e^{-ig \int_{\Gamma_1}\bB dx}e^{-ig \int_{\Gamma_2}\bB dx}\rag_B ~  .
 \enq
  This quark-quark-scattering  amplitude is neither observable nor
gauge invariant. In order to construct a gauge invariant expression
 we consider the  scattering of two colour neutral quark-antiquark
states, so called colour dipoles. This leads to the  expectation
value of two Wegner-Wilson loops~\ct{DFK94}, as depicted in Fig.
\ref{twoloops}.

 The four corners of the loops have the coordinates
 \beqa
C_1&:&(-T,-T,\vec x\,_1),(T,T,\vec x\,_1),(-T,-T,\vec {x'}\,_1),(T,T,\vec {x'}\,_1)\\
C_2&:&(-T,T,\vec x\,_2),(T,-T,\vec x\,_2),(-T,T,\vec {x'}\,_2),(T,-T,\vec {x'}\,_2)
\enqa
where the arrow always indicates vectors in transverse space and $T$
goes to infinity.

The relative and centre coordinates are introduced as
\beq
\vec{R_i}= \vec x\,_i - \vec {x'}\,_i; \quad \vec X_i
= \vec {x'}\,_i + z_i \vec R_i   ~ .
\enq
The vector $\vec R_i$ denotes the transverse extension of the loop $i$
and the quantity  $z_i$ with $0 \leq z_i \leq 1$ will later be identified with
the longitudinal
 momentum fraction of the quark. The impact parameter vector
$\vec{b}$ is defined by
\beq
\vec{b} = \vec{X}_1-\vec{X_2} ~ .
\label{ko2}\enq
With this definition the $t$-dependent scattering
amplitude can be obtained as
the two-dimensional Fourier transform with respect to the impact
parameter~\ct{DGKP97}.

The basic element of the scattering matrix for colour-singlet
quark-antiquark dipoles is  the expectation value of two loops,
\beq
S(\vec{b},\vec R_1,z_1,\vec R_2, z_2))=
\frac{\textstyle{\frac{1}{9}}\lag W[C_1]
W[C_2]\rag}{\textstyle{\frac{1}{3}}\lag W[C_1]\rag
\textstyle{\frac{1}{3}}\lag W[C_2]\rag} \label{smatrix} ~ ,
\enq
with
\beq
W[C_i] = {\rm tr~P}\exp[-i g \oint_{\cS_i} \bB dx]
\label{loop}\enq
where $\cS_i$ the border line of loop $i$.

We will discuss later how we pass from the dipole-dipole  amplitudes
to hadronic scattering (or photoproduction) amplitudes by
integrating over light-cone wave functions.

The expectation value of the loops is  approximately calculated
using an extension of the stochastic vacuum model (SVM).
In this model it is assumed that the long-distance behaviour of QCD
can be approximated by a Gaussian stochastic process with
the gluon field strength as the stochastic variable. This model
yields confinement in non-Abelian gauge theories and is in
conformity with the Mandelstam-t'Hooft~ \cite{Man76,tHo76} picture of
string formation through monopole condensation;
we refer to \cite{DSS00} for a detailed review.
In order to apply the model for the evaluation of two loops it has
to be extended, since in a stochastic process with non-commuting
variables the higher cumulants which must vanish in a Gaussian process
are not uniquely defined.

In order to pass from the gluon potential occuring in the line integrals
in Eq. (\rf{loop})  to the gluon field strength
we have  to apply the non-Abelian Stokes theorem. This implies a special
choice of a surface containing both loops as borders. This choice is
 not unique, and in this paper we use one proposed in \ct{DFK94}; for other
possible choices, see \ct{SSP02}.

The general form of the basic correlator which determines the full
Gaussian process is in our approach given by~\ct{DS88}
\begin{eqnarray} \lefteqn {\langle :g^2
F^a_{\mu\nu}(x) \Big( e^{-ig \int_x^{x'} A dz }\Big)^{bc}\ F^d_{\rho\sigma}(x'):
\rangle=}\nn\\
&&~~~{\textstyle{{1}\over{96}}}\de^{ab} \de^{cd}
\langle g^2 FF\rangle_B \int
{d^4k\over (2\pi)^4}\ e^{-ik.{(x-x')}}
\Big((g_{\mu\rho} g_{\nu\sigma}-
g_{\mu\sigma} g_{\nu \rho})\ \kappa ~ i\widetilde D(k^2) \nonumber \\
&&~~ + (- g_{\nu\sigma}k_\mu k_\rho+g_{\nu\rho}k_\mu
k_\sigma-g_{\mu\rho}k_\nu k_\sigma +g_{\mu\sigma}k_\nu
k_\rho)
(1-\kappa)\ i \frac{d\widetilde {D}_{1}
(k^2)}{dk^2}\Big) ~ .\nn\\
\label{msvc}
\end{eqnarray}
For the correlation functions we take the form proposed in \ct{DFK94}
\beqa
\tilde D(k^2)&=&
\frac{27 \pi^4 k^2}{4 a^2 (k^2+\frac{9 \pi^2}{64 a^2})^4 } ~ ,  \nn \\
\tilde D_1(k^2)&=&
\frac{9\pi^4 }{2 a^2(k^2+\frac{9 \pi^2}{64 a^2})^3 } ~ ,
\label{msv4}\enqa
with the parameters
\beq
a=0.346 ~  {\rm fm}~ ,  \qquad \lag g^2 FF\rag a^4 = 23.5 ~ , \qquad \kappa=0.74 ~ ,
\label{input1} \enq
where $a$ is the correlation length, which are  in agreement  with lattice results~\ct{DGM99,Meg99}.
This same parameter set  has been used for many applications of the
model to hadron-hadron, photon-hadron and photon-photon high energy
interactions and will be used  throughout this paper.

The formalism set up above allows us to calculate the
scattering matrix in terms of the correlator (\rf{msvc}) using
the assumptions of the stochastic-vacuum model.
The most straightforward way is to expand the exponentials occuring
in the Wegner-Wilson loops of Eq. (\ref{smatrix}). We then obtain
\beqa S(\vec{b},1,2)&=&
{\textstyle \frac{1}{144}}\tr(\lam^a\lam^b)\tr(\lam^c\lam^d) \nn \\
&&\hspace{-2cm}\times  \Big \langle \int_{\cS_1} d\si^{\mu\nu}
F^a_{\mu\nu} \,\int_{\cS_2} d\si^{\ka\lam} F^c_{\ka\lam}
\int_{\cS_1} d\si^{\mu\nu} F^b_{\mu\nu}
\,\int_{\cS_2} d\si^{\ka\lam} F^d_{\ka\lam}\Big \rangle _B \nn \\
&&\hspace{-2cm}+ \dots \label{c14.2}
\enqa
where the dots represent products of more than four field-strength
tensors and 1,2 stands for $\vec R_1,z_1,\vec R_2,z_2$.
Inserting the expressions  (\ref{msv4}) and performing the surface
integrals we are finally lead to
\beq
S(\vec{b},1,2)=
1-{\textstyle{{1}\over{9}}} \chi^2(\vec{b},1,2) ~  ,
\label{c14.6b}\enq
with
\beqa
\chi(\vec{b},1,2)&=& \frac{1}{96} \langle g^2 FF\rangle
\Big(
I(\vec x_1,\vec x_2)+I(\vec x'_1,
\vec x'_2)\nn\\
&&~~~~~~~~-I(\vec x_1,\vec x'_2)- I(\vec x'_1,
\vec x_2)\Big).
\label{c14.7}\enqa
Using the special form of the correlators in Eq. (\rf{msv4}) we obtain
\beqa
I(\vec x_1,\vec x_{2}) &=& \half \pi \ka
\int_0^1dv \Big(|v
\vec x_2-\vec x_1|^2
K_2(\lambda^{-1}|v \vec
{r}_2-\vec x_1|)~~~~~~~~~~~~ \nn \\
&&~~~~~~~~~~~~~~+|\vec x_2-v
\vec x_1|^2 K_2(\lambda^{-1}| \vec x_2-v\vec
{x}_1|)\Big) \nn \\
&&+(1-\kappa) \pi \lambda^2
| \vec x_2-\vec x_1|^2
K_3(\lambda^{-1}|\vec x_2-
\vec x_1|)
\enqa
where  $\lambda = (3\pi/8) a ~ $ , and $K_2$ and $K_3$ are modified 
Bessel functions.

A more refined method to treat the two traces  has been developed by 
Berger and Nachtmann \cite{BN99}. The main idea is to interpret the  
product of the two
separate traces in Eq. (\rf{c14.2}) over $3\times 3$ matrices (generators
of $SU(3)$ in the fundamental representation) as one trace
${\rm Tr}_2$  in the product space of the
two fundamental representations of $SU(3)$; that is ${\rm Tr}_2$ acts
in $SU(3)\otimes SU(3)$. Thus
\beqa
S(\vec{b},1,2)&=&
\textstyle{\frac{1}{9}} {\rm Tr}_2 \Big\langle \exp\Big(-i g \int_{\cS_1}
d\si^{\mu\nu}
F^a_{\mu\nu}(\half \lam^a \otimes 1) \Big) \nn \\
&&\times\exp\Big(-i g  \int_{\cS_2} d\si^{\mu\nu}
F^c_{\mu\nu}(1\otimes\half \lam^c \big)\Big)\Big\rangle_B ~.\label{bn}
\enqa
The two exponentials commute in the product space and we can write
the right-hand side of this equation as a single exponential
\beq
S(\vec{b},1,2)=
\textstyle{\frac{1}{9}} \, {\rm Tr}_2 \Big\langle \exp\Big(-i g \int_{\cS}
d\si^{\mu\nu}
\bF_{\mu\nu}\Big) \Big\rangle_B \label{c14.4} ~ ,
\enq
where the surface integral extends over the surfaces $\cS_1$ and $\cS_2$,
 and $\bF_{\mu\nu}$ takes its values in the product algebra of
$SU(3)\otimes SU(3)$. We can now make a cluster expansion
with stochastic variables from the product algebra and
using the correlator (\rf{msv4}) we  finally  obtains~ \cite{BN99}
\beq
S(\vec{b},1,2)=
\textstyle{\frac{2}{3}} e^{- \frac{1}{3}i\chi} + \textstyle{\frac{1}{3}}
e^{\frac{2}{3} i \chi}   ~ ,
\label{c14.6}\enq
where  $\chi=\chi(\vec{b},1,2)$ is the same as  given in Eq. (\rf{c14.7}).

In the following we refer to the first method leading to Eq. (\ref{c14.6b})
as the expansion method, and the second method leading to Eq. (\ref{c14.6}) 
as the matrix-cumulant method.

In a very loose sense we can consider the quantity $\chi$ as
representing the exchange of a nonperturbative
gluon. In this sense the expansion method takes only into
account  the exchange of two nonperturbative gluons.
The matrix cumulant method then  takes  also into account
multiple gluon exchange; it will automatically satisfy unitarity
constraints for hadronic cross sections.
An expansion of Eq. (\rf{c14.6}) in $\chi$ yields in leading order
the result obtained in Eq. (\ref{c14.6b}) above, namely
\beq
S(\vec{b},1,2)= 1-
\textstyle{\frac{1}{9}} \chi^2 - \dots
\lb{expodd}\enq

\begin{figure}[ht]
\vskip 2mm
  \includegraphics[width=12cm]{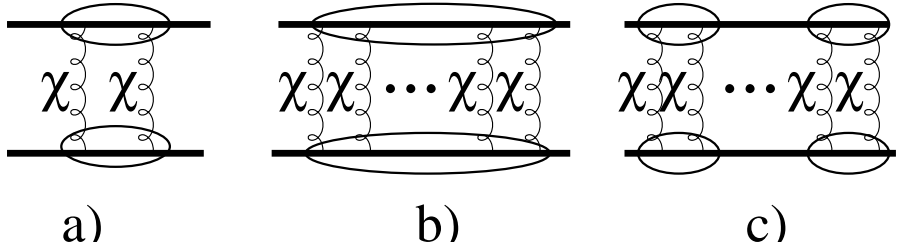}
\caption{\label{methods} Illustration of a) the expansion metod, b) the
matrix-cumulant method and c) the eikonal method. The gluon line with a 
$\chi$ corresponds to the expression (\ref{c14.7}) and can be viewed 
loosely as the exchange of a nonperturbative gluon.}  \end{figure}

We also consider the eikonal unitarisation method
\beq
S(\vec b ,1,2)= e^{-\chi^2/9} ~ , 
\label{eik1}\enq
which is similar to forms used to study saturation effects 
\cite{pqcd4,CFKS01}. 
Like the matrix-cumulant method it corresponds to multiple exchange, but here
the exchanged objects represented by $\chi$ are coupled in such a way that
always a pair forms a color singlet whereas in the matrix-cumulant method only
the entirety of the exchanged objects has to form a colour singlet. The three
different methods are illustrated in Fig. \ref{methods} ~ .

The matrix element for
dipole-dipole scattering with momentum transfer $\vec{q}$ is
then given by
\beq
T_{fi}(s, t, \vec{R}_1,z_1,\vec{R}_2,z_2)= -2 i s  \int d^2 b\,
e^{i \vec
{q}.\vec{b}} \Big( S(\vec{b},1,2)-1\Big) ~ ,
\lb{c14.8}
\enq
with $t=-\vec{q}^2$.

The treatment of three quarks in a colour-singlet state  is analogous
to the scattering of two dipoles,
but technically more involved. We refer to the literature
\cite{DFK94,BN99} for the corresponding results.

If the parameters of  the stochastic-vacuum model are taken to be
independent of the energy, the resulting cross sections turn out to
be independent of the scattering energy too.
The observed energy dependence has therefore to be introduced by hand.
One way is to make the size of the hadrons energy dependent
~\cite{DFK94,FP97}. This has been shown to be very simple and effective
for purely hadronic processes. We here adopt the two-Pomeron approach
of Donnachie and Landshoff \ct{DL98}, coupling a  soft Pomeron
with intercept 1.08 to large and a hard Pomeron with intercept 1.42
to small dipoles \ct{DDR00,DD02}.

Specifically we introduce an energy dependence of the quantity $\chi$,
Eq. (\ref{c14.7})
\beqa
\chi(b,1,2) &\to& \chi^E(b,1,2,s)=\chi(b,1,2) \Bigg(\frac{s}{s_0}\Bigg)^{\epsilon_s/2} \mbox{ for } R_1
\mbox{ and }R_2 >r_c \nn \\
\chi(b,1,2) &\to& \chi^E(b,1,2,s)=\chi(b,1,2) \Bigg(\frac{s R_i^2}{s_0 r_c^2}\Bigg)^{\epsilon_h/2} \mbox{ for }
 R_i\leq r_c, i=1 \mbox{ or } 2\nn \\
\chi(b,1,2) &\to& \chi^E(b,1,2,s)=\chi(b,1,2) \Bigg(\frac{s R_1 R_2}{s_0 r_c^2}\Bigg)^{\epsilon_h/2}
\mbox{ for } R_1 ~  \mbox{\bf{ and }} ~ R_2 \leq r_c \nn \\
\label{chiE} \enqa
with $\epsilon_s=0.08,~ \epsilon_h=0.42$ taken from
\ct{DL98} and   $r_c$=0.22 fm  taken from a treatment of the proton
structure function  \ct{DD02}.

\subsection{Wave functions and hadronic reactions}

We are interested in reaction amplitudes where the external particles
are physical hadrons and photons. We obtain such amplitudes from the
dipole-dipole scattering amplitude  (\rf{c14.8}) by integrating
over all dipoles sizes with the light-cone  wave functions of the
participating particles as weights.

A meson or a photon is here described by a light-cone wave
function $\psi_n(\vec{R}_i,z_i)$ of a quark and an antiquark with
relative transverse coordinates $\vec{R}_i$ and quark longitudinal
momentum fraction  $z_i$. The meson or photon scattering amplitude for
the reaction $a~b \to c~d$ is then obtained from  the dipole-dipole 
scattering amplitude  $T_{fi}(s,t,\vec{R}_1,z_1,\vec{R}_2,z_2)$ 
of Eq. (\rf{c14.8}) by 
 \beq
T_{a b \to c d}(s,t)=-2is \int d^2\vec b\, e^{i \vec p . \vec b}
    \Big(S_{a b \to c d}(\vec b,s)-1\Big) ~ ,
\enq
where
\beqa
S_{a b \to c d}(\vec b,s)&=&
 \int d^2 R_1 \int d^2 R_2 \int_0^1 dz_1 
   \int_0^1 dz_2 ~ \psi_c^*(\vec{R}_1,z_1) \psi_a(\vec{R}_1,z_1)\nn \\
&&\times ~ \psi_d^*(\vec{R}_2,z_2) \psi_b(\vec{R}_2,z_2)S(\vec b, 1,2)~ .
\lb{general} \enqa
The normalization is such that 
 \beq
\frac{d \sigma}{d|t|} = \frac{1}{16 \pi ~  s^2} ~  
                 |T_{a b \to c d}(s,t)| ^2  ~ .
\enq

For the heavy quark content of a photon the perturbative expressions
for the wave functions are reliable since the charm quark mass sets
the scale. Although we need in the present work only with the transverse
wave function, we give for completeness the expressions for both
transverse and longitudinal photons, the latter being needed
for electroproduction of mesons. For photons of helicities 1, -1 and 0,
we write respectively

\begin{eqnarray}
\lefteqn{\psi_{\ga,1}(Q^2;z,r,\theta)=} \nn \\
&&\hat e_f \frac{\sqrt{6\alpha}}{2 \pi} \Big[ i \ep
e^{i\theta}(z\de_{h,+}\de_{\bar h,-}-\bar z \de_{h,-}\de_{\bar h,+})K_1(\ep r)
+m_f\de_{h,+}\de_{\bar h,+}K_0(\ep r)\Big] ~ ,
\end{eqnarray}

\begin{eqnarray}
\lefteqn{
\psi_{\ga,-1}(Q^2;z,r,\theta)=} \nn \\
&&\hat e_f \frac{\sqrt{6\alpha}}{2 \pi} \Big[ i \ep
e^{-i\theta}(\bar z \de_{h,+}\de_{\bar h,-}
-z\de_{h,-}\de_{\bar h,+})K_1(\ep r) +
m_f\de_{h,-}\de_{\bar h,-}K_0(\ep r)\Big]
\label{photontr} \end{eqnarray}
and
\beq
\psi_{\ga,0}(Q^2;z,r)=\hat e_q \frac{\sqrt{3\alpha}}{2 \pi}
 (-2 z \bar z) ~   \de_{h, -\bar h}~ Q~ K_0(\ep r) ~ ,
\label{photonlo}\enq
where
\beq \ep=\sqrt{z\bar z Q^2+m_f^2} ~ ,\enq
and $m_f$ is the quark mass and $\hat e_f$ is the quark charge in
units of the elementary charge for each flavour $f$;
$K_0$, $K_1$ are the modified Bessel functions.

Meson wave functions are more model dependent than photon wave
functions and especially the spin structure can be quite
complicated \cite{BB98}. In this paper we take for the vector
mesons the spin structure from the vector current leading to 
similar expressions as for the photon~\ct{DGKP97}, namely
\beqa
\psi_{V,+1}(z,r) &=& \phi_V(z,r) \Big( i \om^2 r
e^{i\theta}(z\de_{h,+}\de_{\bar h,-}-\bar z \de_{h,-}\de_{\bar h,+}) +
m_f\de_{h,+}\de_{\bar h,+}\Big) ~ , \nn \\
\psi_{V,-1}(z,r) &=& \phi_V(z,r) \Big( i \om^2 r
e^{-i\theta}(\bar z\de_{h,+}\de_{\bar h,-}- z \de_{h,-}\de_{\bar h,+}) +
m_f\de_{h,-}\de_{\bar h,-}\Big)  
\label{mesontr}\enqa
and
\beq
\psi_{V,0}(z,r) =
   \phi(z,r)_V\Big(\om  4 z \bar z \de_{h, -\bar h}\Big) ~ .
\label{mesonlo}\enq
Here $\pm 1$  and $0$  denote transverse and longitudinal polarizations
of the vector meson, and $h$ and $\bar h$ represent the helicities of
quark and antiquark respectively.

The  functions $\phi_V(z,r)$ are constrained by the normalisation
condition and by the electromagnetic decay width, as described in the 
Appendix. Making for the $r$ dependence a Gaussian ansatz,
the parameters are then completely determined  by the two
conditions just mentioned. For the $z$ dependence we make two
ans\"atze: one is suggested by the phenomenologically very successfull
Bauer-Stech-Wirbel model \cite{BSW87}
\beq
\phi_{\rm BSW}(z,r) = \frac{N}{\sqrt{4 \pi~}} ~\sqrt{z(1-z)~}~
\exp\Big[-\frac{M_V^2}{2 \om^2}(z-\frac{1}{2})^2\Big]~~
     \exp[-\frac{1}{2}\om^2 r^2] ~ .
\enq
The other choice is  obtained in the spirit of Brodsky-Lepage~\cite{BL80}
from a non-relativistic wave function in the rest frame where the
transition to the light-cone coordinates
is achieved by the replacement of the relative momentum $\vec k$ of the
two constituents by the transverse momentum $\vec k_T$ and the
longitudinal momentum fraction of the quark according to
\beq
k^2 \to \frac{k_T^2+m_f^2}{4 z (1-z)} - m_f^2   ~ .
\enq
This procedure only makes sense for a finite quark mass and  leads to a more photon-like $z$ dependence
of the wave function. We then write
\beq
\phi_{\rm BL}(z,r) = \frac{N}{\sqrt{4 \pi~}}
\exp\Big[-\frac{m_f^2(z-\frac{1}{2})^2}{ 2 \om^2  z(1-z)}\Big]~\exp[-2
z(1-z)\om^2 r^2] ~ .
\enq
$M_V$ and $m_f$ represent respectively the vector meson mass
and the the quark  mass.

For the proton we should use a three quark wave function which in
principle poses no problems \ct{DFK94}. Earlier investigations have
shown however that a quark diquark structure of the proton leads to
phenomenologically consistent results and most applications have 
been made in this
picture. The quark-diquark nucleon wave function can be treated like
a quark-antiquark meson wave function and again we make a Gaussian
ansatz for the wave function and fix the longitudinal momentum
fraction at $z \approx 1/2$ ~,  namely 
\beq
\psi_p(R)= \frac{1}{2\pi} \frac{1}{S_p} e^{-r^2/(2S_p)^2}  ~ .
\label{proton} \enq
 The transverse size parameter $S_p$
of the proton is choosen to be
\beq
S_p=0.74 ~~~{\rm fm} ~ ,
\label{inrp}\enq
which leads to a good description of $p p$ scattering and also to
a good proton form factor \ct{DNPW01}.

 For the charm mass we adopt  the  value used in a previous analysis
of structure functions~\ct{DD02},
\beq
m_c=1.25 ~ \rm{GeV} ~  .
\label{charmass} \enq
This value lies in the middle of the range
of masses  of the modified minimal subtraction scheme~\cite{PDG02}.
The dependence of the charm mass is discussed in the Appendix.

 This finishes the description of the general model, with 
characterization of all parameters. It should be noted that in 
the original application of   
the matrix cumulant method \ct{BN99}  a slightly different set
 of parameters was used in order to get an optimal overall
description of the elastic proton-proton  cross section. Showing 
in the present  paper the values of amplitudes and cross sections
 evaluated with only one  set of parameters we wish to  emphasize 
the magnitude of the influence of multiple scattering mechanisms.

\subsection{Photoproduction of vector mesons}

Using the results of the two previous sub-sections we can calculate
the scattering amplitudes for photoproduction of vector mesons.

Inserting the wave functions of Eqs.({\ref{photontr}),(\ref{mesontr})
and (\ref{proton})
into the production amplitude (\rf{general}) we obtain
\beq
S_{\gamma p \to V p}(b)= \int d^2 R_1 ~ \int dz_1\, \int d^2 R_2 \, \rh_{\gamma,V,\lambda}(z_1,R_1) |\psi_p(R_2)|^2
S(b,z_1,\vec R_1,1/2,\vec R_2) ~ ,
\label{int} \enq
where $\rh_{\gamma,V,\lambda}(z_1,R_1)$ is the photon-vector-meson
overlap function
\beqa
\rh_{\gamma,V,\lambda}(z_1,R_1) &=& \hat e_q\frac{\sqrt{6\alpha}}{2\pi} 
   ~~\phi_{BSW/BL}(z,R_1)\nn \\
&\times& \Big( \ep ~\om^2 R_1 \big[z^2+(1-z)^2\big] K_1(m_f~R_1)
   + m_f^2 K_0(m_f ~R_1)\Big) ~ .
 \enqa

The overlap densities are independent of the angles
$\theta_1$ and $\theta_2$ of $\vec R_1$ and $ \vec R_2$, and the
functions $\chi(\vec b,1,2)$ change sign
if $\vec R_1$ or $\vec R_2$ is reversed, as can be seen easily from
Eq. (\rf{c14.7}) by noting that $\vec R_i \to -\vec R_i$ corresponds to
$(\vec x_i,\vec x'_i) \to (\vec x'_i,\vec x_i)$.
Inserting the result (\ref{c14.6}) of the matrix cumulant method
into Eq. (\rf{int}) we see  that
after integration over the angles $\theta_1$ and $\theta_2$  only
the even terms in $\chi$ survive and
the exponentials in Eq. (\ref{c14.6}) can be replaced  by cosines.
Therefore we may insert in Eq. (\ref{int}), as result of the matrix
cumulant method, 
\beq
S(\vec b ,1,2)=
\textstyle{\frac{2}{3}} \cos\Big(\frac{1}{3}\chi\Big)
 + \textstyle{\frac{1}{3}}
\cos\Big(\frac{2}{3}  \chi\Big)
  \equiv 1+\frac{2}{3} \Big[\cos\Big(\frac{1}{3}\chi\Big)+2\Big]
   \Big[\cos\Big(\frac{1}{3}\chi\Big)-1\Big]
\label{MC2}\enq

This form guarantees that the
amplitude remains inside the unitarity bounds if $\chi $
becomes large. Differences with respect to the expansion method,
Eq. (\ref{c14.6b}), increase at high energies.

If the energy dependent expressions in Eq. (\ref{chiE})
are introduced it turns out that the hard part of the proton,
$R_1\leq r_c$
gives for energies below the TeV region only
a negligible contribution, so we have only to consider the two
cases $R_1\leq r_c$ and $R_1 > r_c$.

Now all  formulae are set up and all  parameters are fixed.

Our results for the differential and total cross section of $J/\psi$
photoproduction in the expansion method are written 
\beq
\frac{d \sigma}{d|t|} =\bigg( A_{ h}(t) \Big(\frac{s}{s_0}\Big)^{0.42}+
A_{ s}(t) \Big(\frac{s}{s_0}\Big)^{0.08}\bigg)^2  ~ .
\label{para1} \enq
The soft and hard amplitudes for the case of the BSW wave function 
can be conveniently expressed by the  parametrisations
\beqa
A_s(t) &=& \frac{10.82\exp(-1.07 |t|)}{1+3.39|t|} ~ ,  \nn \\
A_h(t) &=& \frac{2.0\exp(-0.94 |t|)}{1+3.39|t|}  ~, ~   (BSW) ~  ,  
\label{para2}
\enqa
where $t$ is in GeV$^2$ and $A_{s,h}(t)$ in $\sqrt{\rm nb}/$GeV,
$s_0=(20~ {\rm GeV})^2$.
For the case of the BL wave function the forms are very similar
\beqa
A_s(t) &=& \frac{11.17\exp(-1.06 |t|)}{1+3.26|t|} ~  ,  \nn \\  
A_h(t) &=& \frac{2.1\exp(-0.95 |t|)}{1+3.26|t|} ~ , ~  (BL) ~ .  
\label{para3}
\enqa

These very convenient parametrisations  reproduce the
exact results with an accuracy always better than 5 percent.
We see that the t-dependence of the differential cross section predicted by
our model is not a pure exponential, showing a curvature in a
logarithimic scale. Since the parameters expressing the t-dependences
of the hard and soft parts of the amplitude have rather similar 
values, the shape
of the angular distribution (and the slope parameter)
depend only weakly  on the energy (we recall that we  work with
small values of $|t|$).
The values of $A_s(t)$  and $A_h(t)$ are shown in Fig. (\ref{ampl}),
in solid line for  the BWS and in dashed line for the BL wave function.
These forms, together with Eq. (\ref{para1}) contain all   
results for the expansion method that will be used for comparison 
with experiments.  At low energies the soft contribution is several
 times stronger than the hard one. The hard part  reaches the soft 
part for $ W$ about 240 GeV.

Since the differences between the two kinds of 
$J/\psi$ wave functions are very small, from now on in the 
present paper, we will use only one of them, namely the BSW 
wave function. 

The results for the matrix cumulant method~\cite{BN99} cannot be
parametrized so easily, due to the nonlinear
 dependence of the amplitudes on the quantity $\chi^E(b,1,2,s)$,
Eq. (\ref{chiE}). That is, we cannot factorize the $s$ and $t$
dependences of the soft and hard parts, as in Eq. (\ref{para1}).
The same is true of the amplitudes obtained with the eikonal
form of Eq. (\ref{eik1}). In Fig. \ref{ampl} we draw also the
amplitudes for these two unitarization procedures, for a fixed
energy $W= 20$ GeV, where we see that the form of the t-dependence
(in a limited $|t|$ interval) is not much affected by the saturation 
corrections. The same is true for  higher energies.

Accurate parametrisations of the differential cross sections can be 
written in the form  
\beq
\frac{d \sigma}{d|t|}= \frac{A \exp(-a |t|)}{1 + b |t|} ~ , 
\lb{parat}
\enq
yielding results summarised in Tables \ref{tdep1},
 \ref{tdep2}. We found that parametric forms with  dipole factor 
like $1/(1+b|t|)^2$ do not lead to equally accurate representations for 
$d{\sigma}/d|t|$.  


\begin{table} 
\begin{center} 
\begin{tabular}{|c|c|c|c|}\hline 
  &exp.&matrix-cum.& eikonal  \\ 
\hline 
$W$ & A            & A        & A       \\ 
GeV & nb/GeV$^2$  
    & nb/GeV$^2$ 
    & nb/GeV$^2$ \\ 
\hline 
20   & 167.6 & 153.7 &  141.1  \\ 
200  & 896.4 & 790.2 &  693.3  \\ 
1000 & 5657   & 3983   & 3008   \\ 
\hline 
\end{tabular} 
\end{center} 
\caption{Result for $A$ in the parametrisation (\ref{parat}) 
 for the expansion,
the  matrix-cumulant and the eikonal methods.} 
\label{tdep1} 
\end{table} 

\begin{table} 
\begin{center} 
\begin{tabular}{|c||c|c||c|c||c|c|}\hline 
  &\multicolumn{2}{c||}{expansion}&\multicolumn{2}{c||}{matrix-cumulant}& 
                                  \multicolumn{2}{c|}{eikonal}  \\ 
\hline 
$W$  &   a  &   b  &  a   & b   &  a     & b   \\ 
GeV & GeV$^{-2}$ & GeV$^{-2}$ 
    &  GeV$^{-2}$ & GeV$^{-2}$ 
    &  GeV$^{-2}$ & GeV$^{-2}$\\ 
\hline 
20    & 2.72 & 9.0 &  2.83 & 10 &  3.00 & 10 \\ 
200   & 2.62 & 9.2 & 2.87 & 10 &  3.15 & 10 \\ 
1000   & 2.56 & 9.3 &  3.38 & 10 &  3.78 & 10 \\ 
\hline 
\end{tabular} 
\end{center} 
\caption{Results for parameters $a$ and $b$  (\ref{parat}) 
for the expansion,
the  matrix-cumulant and the eikonal methods at different energies.} 
\label{tdep2} 
\end{table} 

\begin{figure}[ht]
\vskip 2mm
\includegraphics[width=10cm]{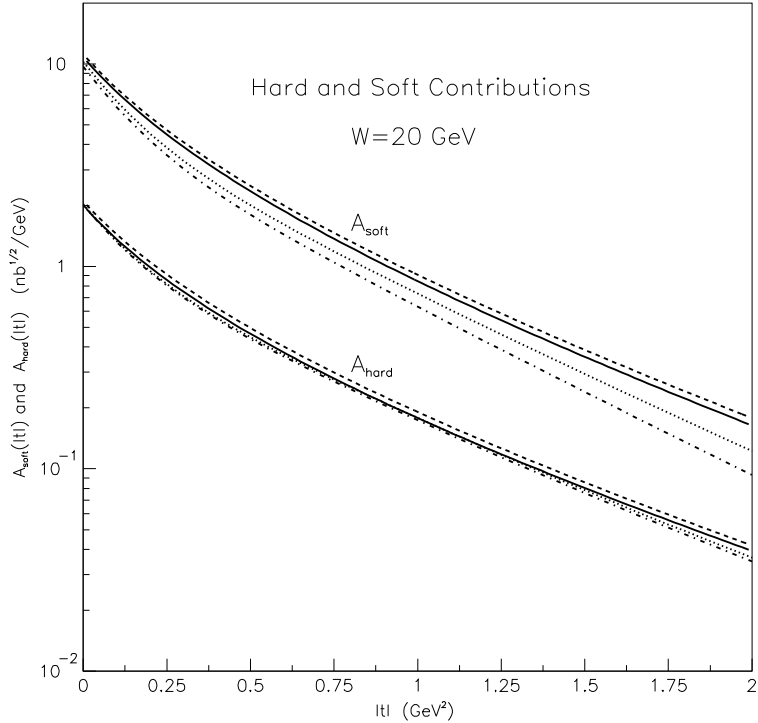}
\caption{\label{ampl} $t$-dependence of soft and hard production amplitudes
at the energy $W=\sqrt{s}=20$ GeV. The lines  can be parametrised in
forms $A(0) \exp(-a|t|)/(1+b|t|)$, as explained in the text.
The solid and the dashed lines are 
the results for the expansion method Eq.(\ref{c14.6b}), using respectively 
BSW and BL wave functions for $J/\psi$. The dotted line corresponds to the 
matrix-cumulant method Eq.(\ref{MC2}) and the dot-dashed line to the  
eikonal method Eq.(\ref{eik1}), both using BSW wave function. }
 \end{figure}

\begin{figure}[ht]
\vskip 2mm
\includegraphics[width=10cm]{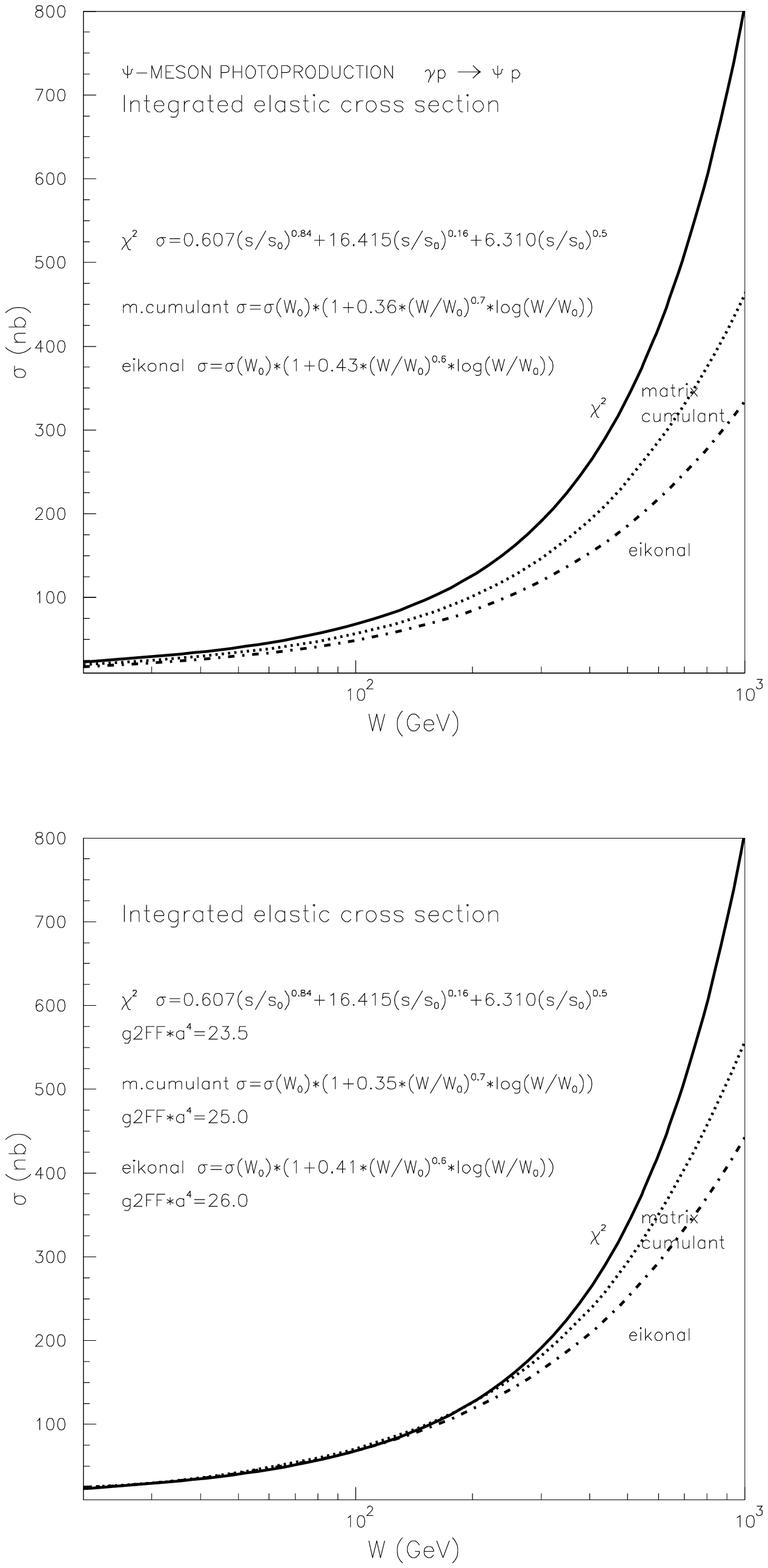}
\caption{\label{tot} Integrated elastic production
$\gamma~  p \to J/\psi~  p $ cross section. In full
line the result with the expansion method, Eq.(\ref{c14.6b}). Dotted
and dash-dotted lines correspond respectively to the matrix cumulant
method of Eq. (\ref{MC2}) and to the exponentiation form of Eq.(\ref{eik1}).
In the upper figure the same $\lag g^2 FF\rag a^4 = 23.5 $ for the
gluon condensate is used for the three curves. In the lower figure we show
that the unitarization effects are compensated, in the present experimental
range $ \sqrt{s} \leq 300$ GeV, by slight changes of parameters:
$\lag g^2 FF\rag a^4=$  25.0 and 26.0 respectively for Eqs. (\ref{MC2})
and (\ref{eik1}). The behaviour at high energies is explained in the
text.}
\end{figure}


For the integrated production cross section we obtain with the
expansion procedure
\beq
\sigma= 0.552 \Big(\frac{s}{s_0}\Big)^{0.84}+
13.539\Big(\frac{s}{s_0}\Big)^{0.16}+5.465\Big(\frac{s}{s_0}\Big)^{0.50} ~ .
\label{para4} \enq

In  the upper part of Fig. \ref{tot}
we show the results for the integrated 
cross sections using the  same input-parameter
set of Eqs. (\ref{input1}) and (\ref{inrp})  for the expansion method
of Eq. (\ref{c14.6b}) and for the  matrix cumulant expression of
Eq. (\ref{MC2}) and for the eikonal  form of Eq. (\ref{eik1}). The
curves show  the influence of possible multiple scattering corrections.
In the lower part of  Fig. \ref{tot}  we instead choose slightly
different values for the gluon condensate:
$\lag g^2 FF\rag a^4=$  25.0 and 26.0 respectively for Eqs. (\ref{MC2})
and (\ref{eik1}). This last figure shows that in the present experimental
range, below 300 GeV, unitarity corrections are easily compensated by
slight changes of one single parameter. Small changes  in the value 
of the charm quark mass that enters in  the wave function  has 
important effects of the same kind. 

We  may conclude that these saturation effects are not clearly observed
in elastic photoproduction of $J/\psi$ for energies up to 300 GeV.

It is interesting to investigate the behaviour of the
integrated cross section at higher energies.  While with the
expansion method the cross section increase according to
Eq. (\ref{para4}), in the unitarized cases
 the behaviour in the whole range  $20 \leq W \leq 1000 $ GeV
is described by the slower behaviour
\beq \sigma(W)=\sigma(W_0) \times
           [1+C \big(\frac {W}{W_0}\big)^\delta
          \log \big(\frac {W}{W_0}\big) ]   ~ ,
\label{tot-uni} \enq
with
 $\delta$ = 0.7 for the matrix cumulant (dotted lines) cases
and = 0.6 for the exponentiation procedure (dot-dashed lines).
 The values of the multiplicative constant C vary a little
with the choice of  $\lag g^2 FF\rag a^4=$ as follows: C=0.36
and 0.35 in the matrix cumulant method respectively for 23.5
and 25.0 , and C=0.43 and 0.41  in the exponentiation procedure
respectively for gluon condensates 23.5 and 26.0.

 If we restrict ourselves to the experimental energy range from
20 to 300 GeV, the results obtained  with the expansion method can be 
represented with an effective  power like form 
\beq \sigma(W)=14.6+9.37 \big(\frac{W}{W_0}\big)^{1.08} ~ .  
\label{range} \enq


The functional integrals occuring in the basic expression
(\ref{smatrix}) for the S-matrix are in our approach
approximately evaluated with the help of the stochastic vacuum
model. This model is an approximation appropriate only for the
soft part of QCD.
The higher the momentum transfer the more important will become
contributions of hard gluons. Therefore we expect our calculations to
be best at small momentum transfer. Of course we cannot predict the
exact scale where hard scattering becomes important but investigations
of the gluon distributions in hadrons and virtual photons~\cite{SSDP02}
 indicate that the hard component becomes as important as the soft
part at a transverse momentum of the gluons $|k_T| \approx 1$ GeV,
so that a safe limit for soft process is
$\sqrt{|t|} \ll 1$ GeV. We also expect that for photoproduction
of $\Upsilon$-mesons even at
small momentum transfer hard gluons play a more important role .
There the small size of the $\Upsilon$  suppresses the contribution of
the soft (nonperturbative) gluons as compared to the hard ones.

Another limitation of our approach is given by the WKB approximation
underlying the nonperturbative approach~\cite{Nac91}
 to scattering. Here the path is assumed to be the classical one,
being nearly a straight line. This also
determines the momentum transfer below which the model can be
safely applied.
The WKB approximation should be more reliable
 at high energies, and this restricts the application of the model
to energies higher than  $\sqrt{s}\approx 20$ GeV.

 Apart from the approximations characteristic of the stochastic vacuum model,
 also the wave functions are treated in a very
 simplified form. From previous experience we expect the reliabillity
 of the model to be about 10 \% in the amplitudes.
 Thus, putting all this together, discrepancies between theory and
 experiment  below 20 \% for the production cross sections could
 be considered as natural.

\section{Comparison with experiment}


\begin{figure}[ht]
\vskip 2mm
 \includegraphics [width=10cm]{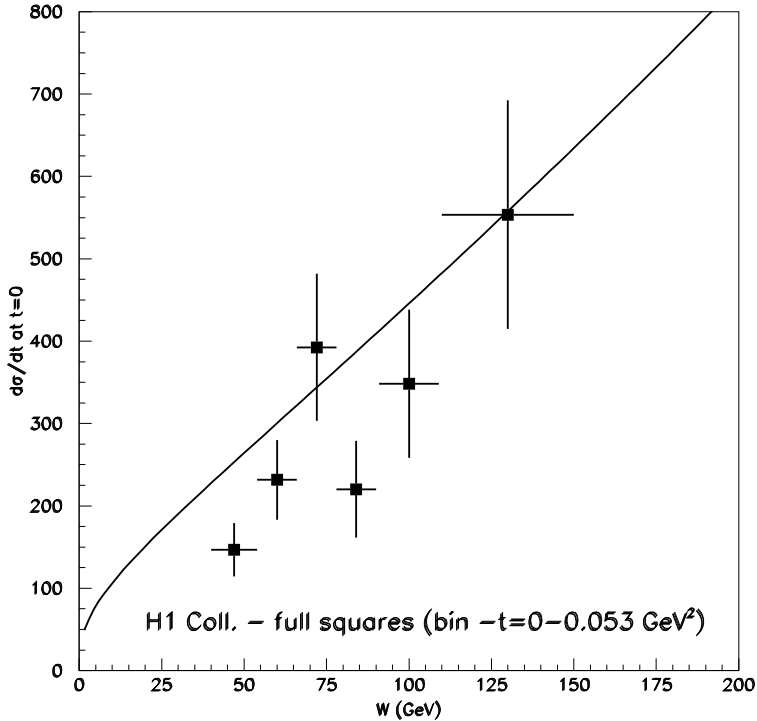}
 \caption{Forward differential cross section $ \big(d\sigma/dt\big)(t=0)$
for the reaction $\gamma~  p \to J/\psi~  p$.
The line represents our results with the expansion method.
The experimental results are from the H1 Collaboration \cite{H1-00}
for the momentum transfer in the interval $0\leq |t|=0.053$ GeV$^2$}
\label{tzero} \end{figure}

 The experimental \cite{H1-00} forward production cross section 
 $ \big(d\sigma/dt\big)(t=0)$
 is shown in  Fig. \ref{tzero}. Only the H1 Collaboration  measures
 directly  this quantity, represented by their
  $0\leq |t| \leq 0.053 $ GeV$^2$  bin . The  values reported by Zeus
\cite{Zeus02} 
 correspond to the extrapolations to $|t|=0$ of their fitted straight
 lines, as will be discussed later (Fig. \ref{tt}).

  A direct comparison with both ZEUS and H1 data in the forward direction
can be made at $|t|=0.1$ GeV$^2$, and this is presented in Fig. \ref{t-one}.
Here we see a very satisfactory agreement between theory  and ZEUS and H1
data, with a remarkable theoretical description of the energy dependence
in the whole range from 40 to 260 GeV.

\begin{figure}[ht]
\vskip 2mm
\includegraphics[width=10cm]{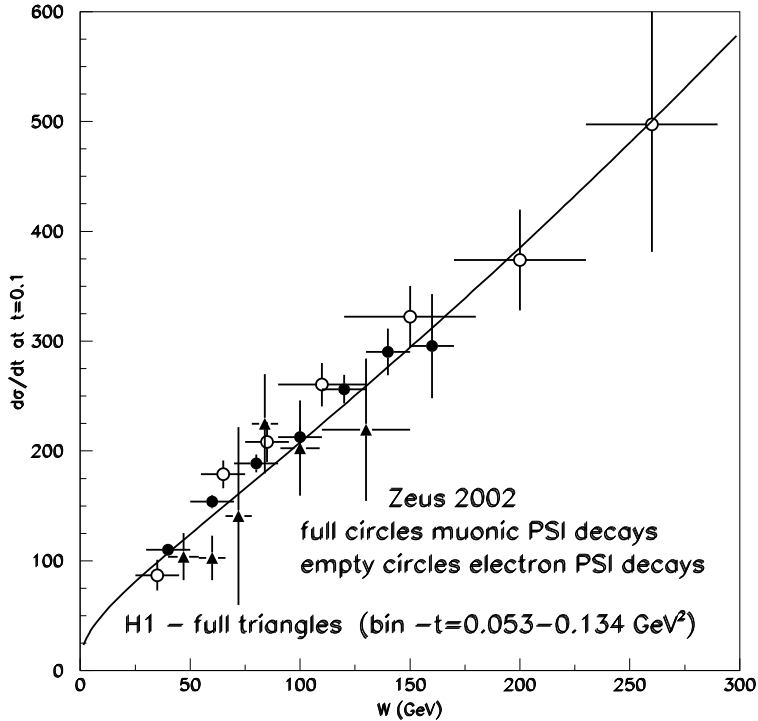}
\caption{Differential cross section  $ \big(d\sigma/dt\big)(|t|=0.1)$
 for the elastic production
$\gamma~  p \to J/\psi~  p$ at $|t| = 0.1$ GeV$^2$.
  The solid line represents our calculation with the expansion method.
The experimental results are from ZEUS \cite{Zeus02} and H1 \cite{H1-00}
collaborations. Zeus data are separately reported for observations
of $J/\psi \to \mu^+ \mu^-$   and  $J/\psi \to e^+ e^-$  decays.}
\label{t-one} \end{figure}

\begin{figure}
\includegraphics[width=12cm]{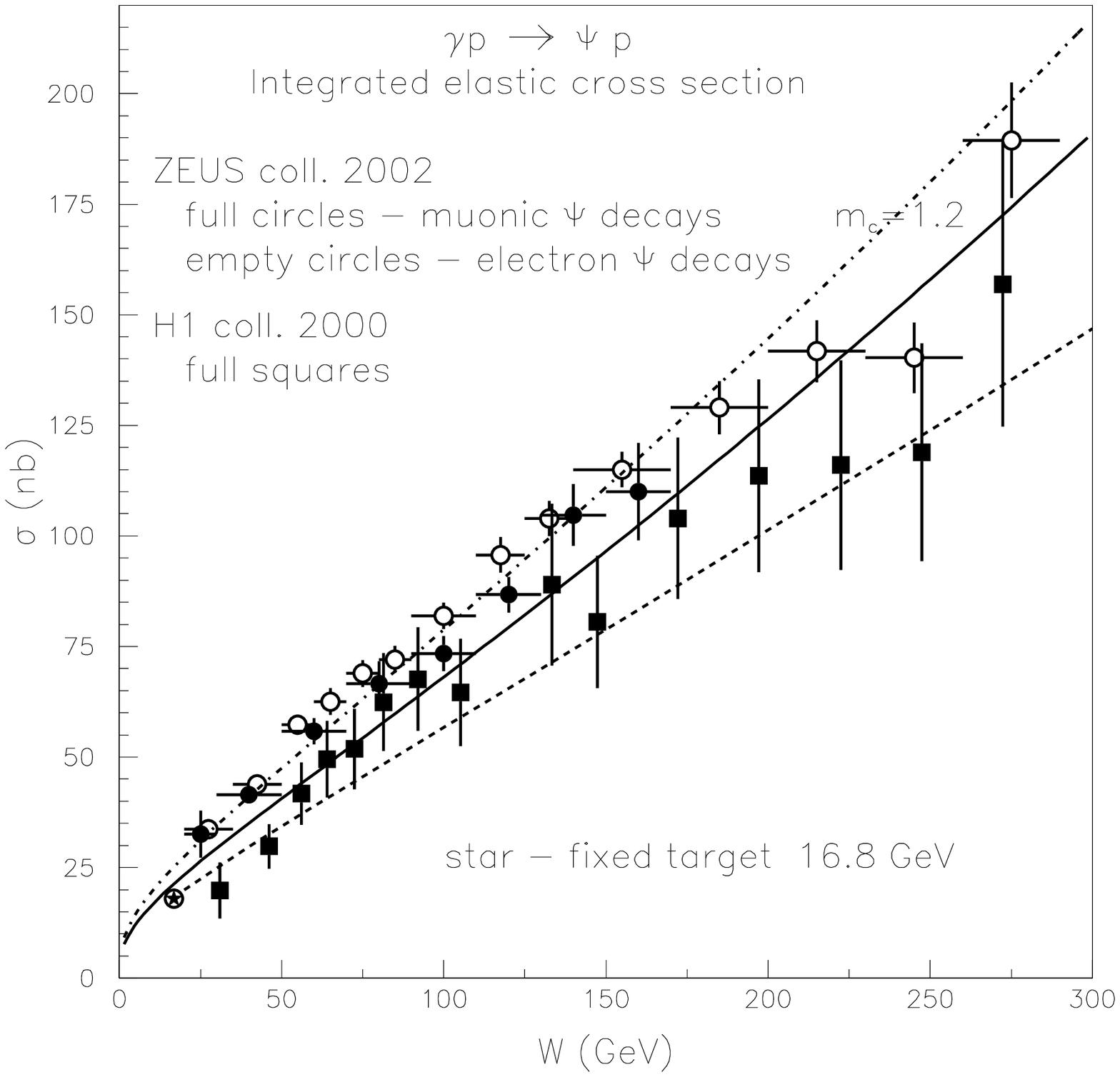}
\caption{Integrated cross section for the reaction
$\gamma~  p \to J/\psi~  p $. The experimental data are from the
H1  \cite{H1-00} and Zeus \cite{Zeus02} collaborations. The
point marked with a star at $W=16.8$ GeV represents  results
from fixed target experiments \cite{target1,target2,target3}.
The solid line is obtained  with the expansion method.
The dashed line shows  our results using the matrix cumulant method,
with the same set of parameters, as in the top part of Fig. \ref{tot}.
To show the influence of a different choice for the charm mass,
the dot-dashed line gives the results obtained with the expansion
method, using $m_c=$ 1.2 GeV. }
\label{resint} \end{figure}

  The integrated production cross section is presented in
  Fig. \ref{resint}, showing again  a reasonable agreement of our model with
  all data up to nearly  300 GeV. The solid line corresponds to  our
 calculation  using the $\chi^2$ expansion of Eq. (\ref{c14.6b}).
  As we have already shown  in Fig. \ref{tot}, the effects of saturation
  introduced by the matrix cumulant or exponentiation method are
  not very large, are within our expected errors, and can be
  compensated by  slight change in parameters. In the figure the dashed
  line represents the results obtained with the matrix cumulant,
  using the standard set of parameters. We have shown
  in Fig. \ref{tot}   that choosing $\lag g^2 FF\rag a^4=$ 25.0 ,
  instead of 23.5, the result of the matrix-cumulant  coincides in this 
   energy range  nearly  with the solid line in Fig. \ref{resint}. To 
   show the influence of the value of the charm mass in the calculation, 
   this figure includes also (dot-dashed line) the results obtained 
  using a value $m_c=1.2$ GeV in the wave functions. 

    The H1 and ZEUS data on integrated elastic cross sections put 
together can be  fitted by either of the two forms (in nb)
\beq
\sigma(W) = 25.8 ~  \big(\frac{W}{W_0}\big)^{0.71}  ~ , 
\label{fit1}
\enq
or  
\beq
\sigma(W) = 26.0+17.14 ~  \big(\frac{W}{W_0}\big)^{0.43} ~  
             \log \big(\frac {W}{W_0}\big)    ~ , 
\label{fit2}
\enq
with the same deviation $\chi ^2= 1.73$ per degree of freedom. 


In Fig. \ref{tt}  we compare our calculations of differential cross sections
with the most recent data from H1 and ZEUS experiments. The energy range is
from 40 to 260 GeV, distributed in 9 bins, and the $|t|$ range is
$ (0-1.6) \rm{GeV}^2$. Where data from
different experiments in similar energy bins are available, we 
plot them together in the same figure. We  emphasize that our
calculation contains no adjustable parameters and that these 
measurements cover a wide energy range.

\begin{figure}[ht]
\vskip 2mm
\includegraphics[width=16cm]{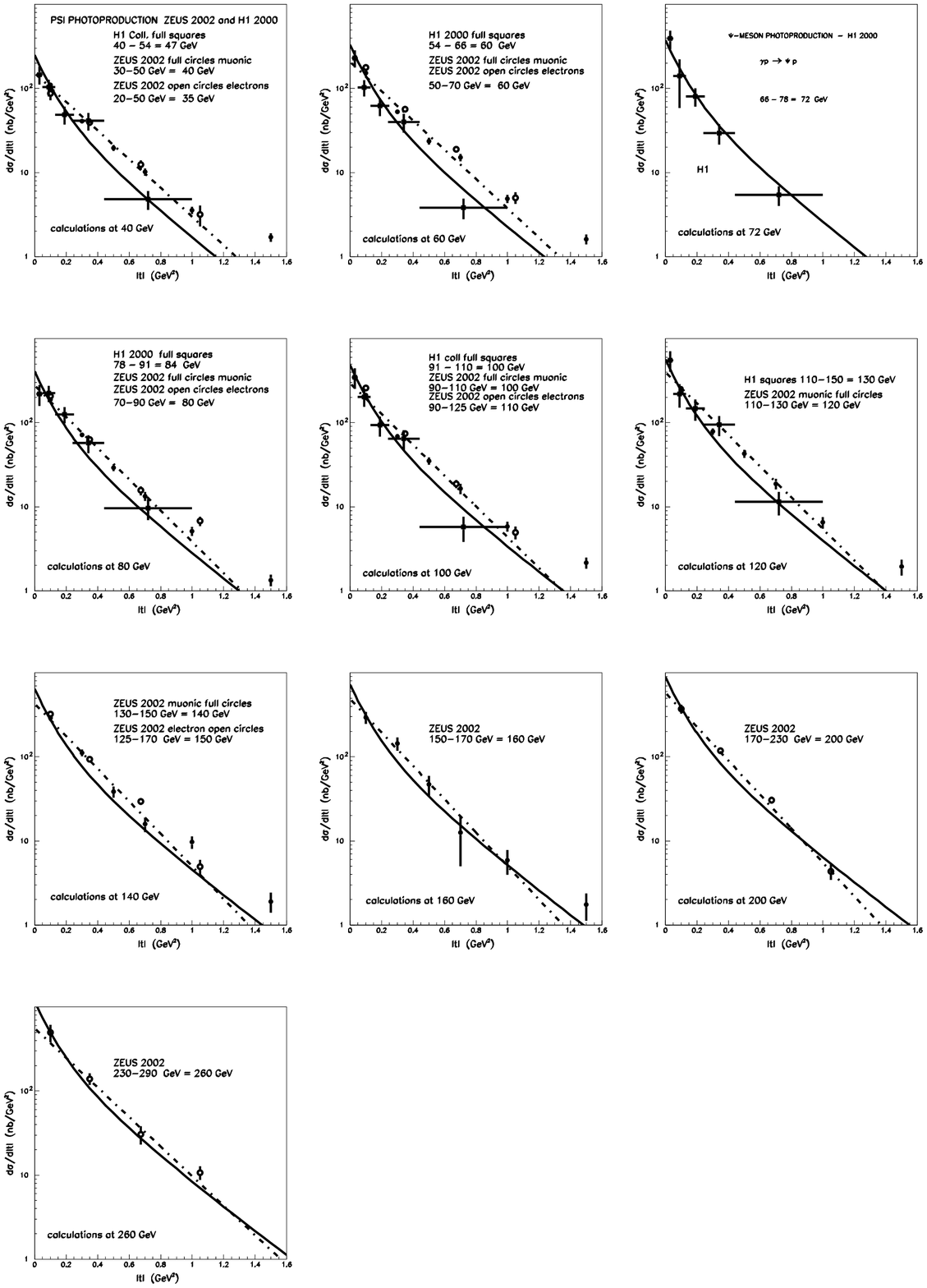}
\caption{\label{tt} $t$-dependence of differential cross sections
for the reaction $\gamma~  p \to J/\psi~  p$. Data are from H1 \cite{H1-00}
and ZEUS \cite{Zeus02} experiments. The solid curves represent our
calculations, as described in the text. The dashed straight lines
are   Zeus fits  in each of their energy bins.}
\end{figure}

 As a quite general feature we see that for small values of the
 momentum transfer $t$ the agreement between experiment and theory
is quite satifactory for all energies.
 For larger values of $|t|$ agreement with H1 data is still
satisfactory, but at the lower and middle energies our results are
in general below the new ZEUS data.
At the highest energies, the agreement improves.
Since in our model we take into account only  nonperturbative effects
it is not surprising that at larger momentum transfers where harder gluons
are  exchanged, some contributions are missing in our calculation.
In \cite{FP02} a good fit up to $|t| \approx 6$ GeV$^2$ is indeed obtained. 

Zeus gives exponential fits in $t$ to their data,  which
 are indicated by  dotted lines in our plots in Fig. (\ref{tt}) .
 From this fit the Zeus collaboration obtains the forward differential
cross section and then evaluates the integrated elastic cross section.
The plots show  remarkable differences  at $|t|=0$ between our
results (with the epansion method) and the ZEUS extrapolated values 
obtained with a straight line.

As we have shown in the general description of the calculation, the
differences between the expansion method and the matrix cumulant
and exponentiation methods are not very large and could be absorbed in
a slightly different choice of parameters. We do not include the
corresponding lines in Fig. \ref{tt} in order not to overload the
plots.

   The data presented above are the most recent HERA data  on $J/\psi$
 photoproduction. Some of the pioneering fixed target experiments of
 20 years ago were made at energies near $\sqrt{s}=20$ GeV , which
 is at the border of the range appropriate for our calculations. We
 have included in
 Fig. \ref{resint} a point at about 16.8 GeV representing this 
 experimental effort  \cite{target1,target2,target3},
 showing that it fits well in the sequence of higher energy data
  points.  As a historical tribute, we show in Fig. \ref{psi17}
 the $|t|$-dependence  of the differential cross sections obtained
  in some of these experiments. In this figure we draw  together 
  our curves for
  the expansion method of  Eq. (\ref{c14.6b}) (solid line),
   for the matrix cumulant result  of Berger and Nachtmann \cite{BN99}
   given by Eq. (\ref{MC2}) and for the exponentiation procedure of
   Eq. (\ref{eik1}).

\begin{figure}[ht]
\vskip 2mm
\includegraphics[width=10cm]{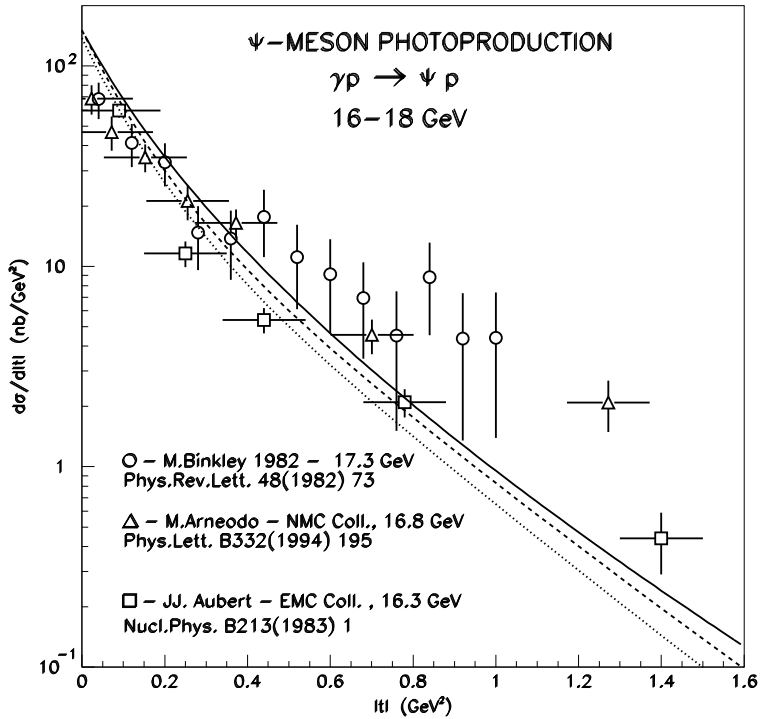}
\caption{\label{psi17} $t$-dependence of differential cross sections
for the reaction $\gamma~  p \to J/\psi~  p$ at the low energies of some
fixed target experiments \cite{target1, target2,target3}. The solid,
dashed and dotted lines represent our results using Eqs.
(\ref{c14.6b}), (\ref{MC2}) and  Eq. (\ref{eik1}) respectively. }
\end{figure}

 We have also applied the model to photoproduction of $\Upsilon~(1S)$
mesons. As expected the results of our  calculation are below the
central values of the experimental data from   Zeus \cite{upsil1}
and H1 ~ \cite{H1-00},  but the
large errors do not allow a definite conclusion about the need 
of other contributions. 
 We find it nevertheless quite astonishing that the
nonperturbative model seems to yield at least a substantial
 fraction of the cross section for such a hard process.
Our results shown in Fig. \ref{upsilon} can be compared with
perturbative calculations
\cite{upsil2}, \cite{upsil3} of the same process.
In \cite{upsil2} the LO calculation taking into account the skewed 
parton distrubution in some way obtains results on the lower edge of 
the error bars, while in \cite{upsil3} additional contributions lead 
to values just through the central values of the measurements.

  \begin{figure}
\includegraphics [width=10cm]{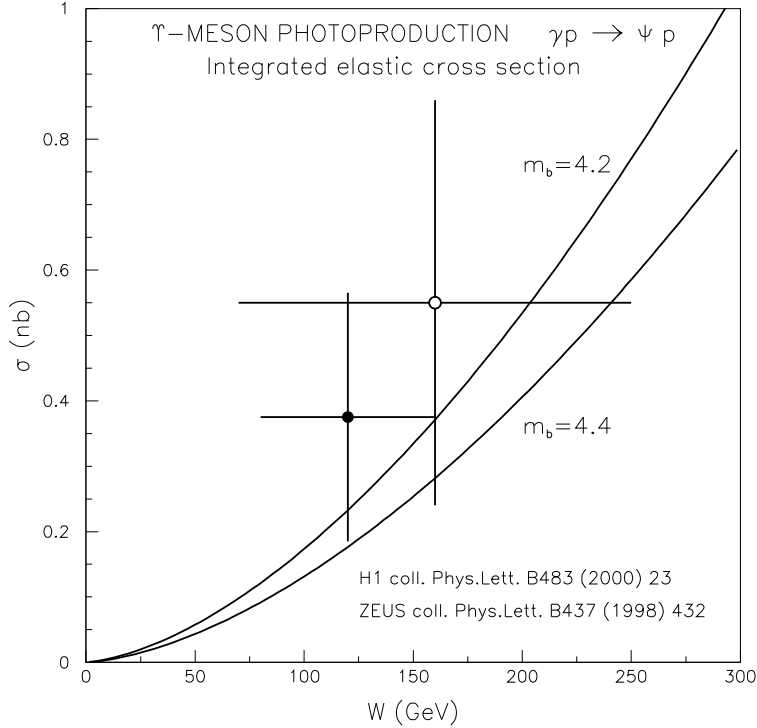}
 \caption{Integrated cross section for the reaction
$\gamma ~  p \to \Upsilon ~  p$. Solid : our result with $m_b=4.2$ GeV,
 dashed  with $m_b=4.4$ GeV. Experimental points from 
Zeus ~ \cite{upsil1} and  H1 ~ \cite{H1-00} }
 \label{upsilon}
 \end{figure}


 \section{Summary and discussion}

The main purpose  of this paper is  to investigate to what extent
photoproduction of $J/\psi (1S)$ mesons can be described in a
purely nonperturbative QCD model. The model has been tested before, 
and  the parameters used in the present work  have been
taken from previous publications on different processes.
They were determined mainly  from results of QCD lattice calculations
and hadronic properties. Our parameter free results are compared
with the more recent experimental  data obtained  at HERA by the
H1   \cite{H1-00} and Zeus  \cite{Zeus02} collaborations.
Comparison of the calculated values of the differential cross sections
with  the data in  nearly forward directions is presented in
Figs. \ref{tzero} and \ref{t-one}. The values reported by Zeus
at $|t|=0$ are not direct measurements, but rather extrapoled
values from linear fits, as shown in Fig. \ref{tt}. In our view,
these linear extrapolations  do not account for possible
structure in the very forward direction, and may lead to
underestimated values, so that Fig. \ref{tzero} only shows the
H1 data. At $|t|=0.1 ~  {\rm GeV}^2$ both H1 and Zeus direct
measurements exist, and Fig. \ref{t-one} shows that the
theoretical model gives excellent description of the
magnitude and of the energy dependence at low $|t|$.

 Fig. \ref{tt} shows that at larger $|t|$ our calculations agree
well with the H1 measurements and are below the Zeus points
for energies up to $W=120$  GeV. At the highest energies the
agreement with Zeus data is good, except that our model
predicts a curvature, with a rise in the very forward direction. 
A clarification of this point is very important, since many models
can only calculate the forward scattering amplitude and
the test of the corresponding theoretical ideas depend
on this kind of experimental data.
Our  model leads to a peculiar form-factor dependence
which is exhibited in Fig. \ref{ampl} and represented by
the parametrizations of Eqs. (\ref{para1}), (\ref{para2})
and (\ref{para3}). The t-dependence of the differential 
cross sections shows  similar curvatures, represented 
by Eq. (\ref{parat}).  

  We recall that our model in meant to be tested in the low $|t|$
range, say below 1.0 GeV$^2$. For larger momentum transfers
 the differential cross section values are already 100 times
smaller than in the soft region,  and an additional
genuinely  hard contribution, however small (without
influence on the integrated cross section), may  play here
an important role.

The integrated cross section is in reasonable agreement with
the experimental data, as shown in Fig. \ref{resint}. 

We have  studied particularly the influence of saturation. We 
have several  methods based on the stochastic  vacuum model to 
evaluate our  model. 
\begin{itemize}
\item The expansion method which
corresponds 
loosely speaking to an  exchange of two ,,nonperturbative'' gluons.
\item The matrix
 cumulant method~\cite{BN99} which takes in some way
multigluon
 exchange into account and respects the unitarity constraints for
hadronic cross sections. 
\item The usual exponentiation  of the profile
function, as in Eq. (\ref{eik1}). 
\end{itemize}
The methods are illustrated in Fig. \ref{methods} and compared  
in Figs. \ref{ampl} and \ref{tot}.
The energy dependence was introduced
based on the two-pomeron model of Donnachie and
Landshoff~\ct{DL98}. In the expansion method it leads to a power
like increase of the integrated cross section, while the matrix
cumulant and in the exponentiation methods there is saturation
due to the unitarity constraints on the dipole cross sections, 
inherently respected by the approach. Up to energies about
1000 GeV, the energy dependence in these unitarity controlled
calculations is very well parametrized in forms given by
Eq. (\ref{tot-uni}).   In Fig. \ref{resint} our calculations
are compared  with the data.

 It should be noted  that in photoproduction processes the Froissart
 theorem cannot be proved and that a powerlike increase does not
 contradict fundamental theorems of local quantum field theory.

At present energies and with present accuracies the data show no
indication of saturation and are well compatible with the
powerlike behaviour of the unconstrained two pomeron approach.
But a comparison with the theoretical approaches that
introduce saturation effects shows that  these are also
compatible with the experiment, considered some 20 \%
allowed variation in our model. Rather small changes of
parameter values may account for differences among calculation
procedures.

We have also tested two different types of wave functions and 
found only small differences in relevant results.

We conclude that, in the energy
range available at HERA, the methods here studied  yield  results
compatible with experiment. As can be seen from Fig. \ref{tot}, 
sizeable deviations are only expected for  energies above 300 GeV.

In our model the residues of both soft and hard pomerons are
 evaluated by nonperturbative methods but this  seems to agree
fairly well with the phenomenological value. This has already been
noted in \cite{DD01} for a large number of total cross sections
and forward amplitudes.

We have extended the use of our model to photoproduction of
$\Upsilon (1S)$ mesons and found that it can yield at least a
sizeable part of this supposedly hard process.

\begin{acknowledgements} Both authors wish to thank DAAD (Germany), 
CNPq (Brazil) and FAPERJ (Brazil) for support of the scientific 
collaboration program between Heidelberg and Rio de Janeiro groups
workin on hadronic physics. The authors are very grateful to Uri 
Maor for discussions and joint efforts for the treatment of $J/\psi$
photoproduction. 
\end{acknowledgements} 

\appendix*

\section{Conditions on wave functions}

The two parameters in the wave functions of the vector mesons,
$N$ ans $\omega$, are
determined by the normalisation condition and the leptonic decay width.

The square of the wave functions (\ref{mesontr},\ref{mesonlo})
summed over internal helicities is given for the
transverse case   by
\beq
|\psi_{X,\pm 1}(z,r)|^2 = |\phi_{X}(z,r)|^2
\times\Big(\om^4 r^2\big[z^2+(1-z)^2\big]+m_f^2\Big)
\enq
and
for the longitudinal case by
\beq
|\psi_{X,0}(z,r)|^2= 2 ~  |\phi_{X}(z,r)|^2 \times (\om 4  z \bar z)^2 ~ .
\enq
where the $X$ in the index stands for BSW ~ \cite{BSW87} or BL ~  \cite{BL80}
-type wave functions.
It has to fulfil  the normalisation condition
\beq
\int_0^1  dz\int|\psi_{X,\lambda}(z,r)|^2 d^2 {\mathbf r} = 1 ~ .
\label{norm}
\enq

The condition which relates the wave function with the e.m. decay width $f_V$
is, for the transverse cases,
\beq
f_V= \hat e_V \frac{\sqrt{6~}}{M_V} \frac{\sqrt{4 \pi~}}{16 \pi^3} ~
  \int_0^1dz\int d^2{\mathbf k} \frac{1}{z(1-z)} \Big(\big[z^2+(1-z)^2 \big]
   k^2+m_f^2\Big)\tilde \phi_{X}(z,k)
\enq
and for the longitudinal cases
\beq
f_V= \hat e_V \om \sqrt{3~} \frac{\sqrt{4 \pi~} }{16 \pi^3} ~ \int_0^1dz\int
d^2{\mathbf k} ~16~ z (1-z) ~\tilde \phi_{X}(z,k) ~ .
\enq
where $\tilde \phi_{X}(z,k)$ is the Fourier transform of
$\phi_{X}(z,r)$, defined through
\beq
\phi_{X}(z,r) = \int \frac{d^2 {\mathbf k}}{4 \pi^2}~
   \tilde \phi_{X}(z,k)~ \exp[-i {\mathbf k} \cdot {\mathbf r}] ~ ,
\enq
and
$\hat e_f $ is the quark charge in units of the elementary charge, that is $\hat e = 2/3$ for the
$J/\psi$ and -1/3 for the $\Upsilon$.

The decay constant $f_V$ is related to the e.m. decay width $\Ga_{e^+e^-}$ through
\beq
f_V^2 = \frac{3 M_V \Ga_{e^+e^-} }{4 \pi \al^2} ~ .
\enq

\begin{table}
\begin{center}
\begin{tabular}{|l|c|c|c|c|c|c|c|}\hline
\multicolumn{1}{|c|}{ } &\multicolumn{1}{|c|}{ }&\multicolumn{3}{|c|}{BSW}&\multicolumn{3}{|c|}{BL} \\
\cline{3-8}
Meson&$m_f$&$\om$&$N$ &$S$& $\om$&$N$&$S$\\
&[GeV]&[GeV]&&[fm]&[GeV]&&[fm]\\
\hline
$J/\psi(1S)  $&$1.2 $&$0.59 $&$3.30$&$0.35$&$0.64 $&$1.47$&$0.36$\\
$            $&$1.25$&$0.58 $&$3.17$&$0.36$&$0.63 $&$1.44$&$0.35$\\
$            $&$1.3 $&$0.57 $&$3.04$&$0.36$&$0.62 $&$1.41$&$0.35$\\
$\Upsilon(1S)$&$4.2 $&$1.29 $&$2.48$&$0.16$&$1.32$&$1.17$&$0.16$\\
$            $&$4.4 $&$1.26 $&$2.35$&$0.16$&$1.30$&$1.14$&$0.16$\\
\hline
\end{tabular}
\end{center}
\caption{ Values of $\om$ , $N$ and size parameter $S$  for transverse
wave functions of $J/\psi$ and $\Upsilon$ mesons.}
\label{mesons2}\end{table}

In Table \ref{mesons2} we give the values of $N,~\omega$ and the
mean square transverse radius for the $J/\psi$ and the $\Upsilon$
wave functions for different choices of the charm and bottom masses
for the transverse BSW- and BL-type meson wave functions.



The $\gamma^*$-vector meson overlap functions, necessary for the
calculation of photo- and  electroproduction of vector mesons are
obtained from Eqs. (\ref{mesontr},\ref{mesonlo}) and (\ref{photontr},
\ref{photonlo}), and are given by

 a) transverse
\beqa
\rho_{\ga^* V,1}(z,r)&=& \hat e_V\frac{\sqrt{6\alpha}}{2\pi} ~~\phi_{X}(z,r)\nn \\
&\times& \Big( \ep ~\om^2 r \big[z^2+(1-z)^2\big] K_1(\ep ~r)
   + m_f^2 K_0(\ep ~r)\Big) =\hat e_V ~ \hat \rho_{\ga V,1}(z,r)
\enqa
b)longitudinal
\beqa
\rho_{\ga^* V,0}(z,r) = -16 \hat e_V\frac{\sqrt{3\alpha}}{2\pi}~ \om~
   \phi_{X}(z,r) z^2 (1-z)^2~Q~ K_0(\ep ~r)
      = \hat e_V ~ \hat \rho_{\ga V, 0 }(z,r)~,
\enqa
where $\ep=\sqrt{z\bar z Q^2+m_f^2}$ ~ .


The real photon-vector meson overlap is obtained by setting $Q=0$, i.e.
the longitudinal part vanishes and $\ep \to m_f$.

\bigskip

The integrals of the square radius over the  overlap functions
\beq
 C(\lambda)=
   \int_0^1 dz \int d^2{\mathbf r} ~ r^2 ~ \rho_{\ga V,\lambda }(z,r)~ .
\label{strength}
\enq
give an estimate of the effective strength for the production
process, the square of these quantities being  nearly proportional
to the integrated cross sections. Thus the numbers in
Table   \ref{overlap} 
tell us that the ratio between $\Upsilon$  and
$J/\psi$ production cross sections at 20 GeV is about $10^{-3}$.

  \begin{table}
  \begin{center}
  \begin{tabular}{|l|c|c|c|}\hline
  \multicolumn{1}{|c|}{ } &\multicolumn{1}{|c|}{ }&\multicolumn{1}{|c|}{BSW}&\multicolumn{1}{|c|}{BL} \\
  \cline{3-4}
  Meson&$m_f$&$C(\lambda=\pm 1)$&$C(\lambda=\pm 1)$ \\
               &[GeV] & [GeV$^{-2}$]  & [GeV$^{-2}$] \\
 \hline
  $J/\psi(1S)  $&$1.2 $&$0.0103   $&$0.0113$\\
  $            $&$1.25$&$0.0095   $&$0.0102$\\
  $            $&$1.3 $&$0.0088   $&$0.0093$\\
  $\Upsilon(1S)$&$4.2 $&$-0.00034 $&$-0.00035$\\
  $            $&$4.4 $&$-0.00030 $&$-0.00031$\\
 \hline
 \end{tabular}
 \end{center}
 \caption{ Values of the integrated square radius over the overlap
 functions, defined by   Eq. (\ref{strength}), for the two kinds of
  transverse wave function.}
  \label{overlap}\end{table}


\begin{thebibliography}{99}

\bibitem{pqcd1}
M.G. Ryskin,
Z.\ Phys.\ C {\bf 57} (1993) 89.

\bibitem{pqcd2}
S.J. Brodsky et al.,
Phys.\ Rev.\ D {\bf 50} (1994) 3134.

\bibitem{pqcd3}
M.G. Ryskin, R.G. Roberts, A.D. Martin, E.M. Levin,
Z.\ Phys. \ C {\bf 76} (1997) 231.

\bibitem{pqcd4}
E. Gotsman, E. Ferreira, E. Levin, U. Maor and E. Naftali,
Phys.\ Lett.\ B {\bf 503} (2001) 277.

\bibitem{pqcd5}
E. Gotsman, E. Levin, U. Maor and E. Naftali,
Phys.\ Lett.\ B {\bf 532} (2002) 37.

\bibitem{Nac91}
O.~Nachtmann,
Annals Phys.\  {\bf 209} (1991) 436.

\bibitem{Dos87}
H.~G.~Dosch,
Phys.\ Lett.\ B {\bf 190} (1987) 177.

\bibitem{DS88}
H.~G.~Dosch and Y.~A.~Simonov,
Phys.\ Lett.\ B {\bf 205} (1988) 339.

\bibitem{DFK94}
H.G. Dosch, E. Ferreira and A. Kramer,
Phys.\ Rev.\ D {\bf 50} (1994) 1992

\bibitem{Nac96}
O.~Nachtmann, Lectures at Schladming School 1996,
 arXiv:hep-ph/9609365.

\bibitem{Dos96}
H.~G.~Dosch,
 ``Nonperturbative Methods in QCD'', in
Hadron Physics 96, Ed. E. Ferreira et al., World Scientific,
     Singapore 1997

\bibitem{Dos99}
H.~G.~Dosch,
Acta Phys.\ Polon.\ B {\bf 30} (1999) 3813.

\bibitem{DGKP97}
H.~G.~Dosch, T.~Gousset, G.~Kulzinger and H.~J.~Pirner,
Phys.\ Rev.\ D {\bf 55} (1997) 2602

\bibitem{Man76}
S.~ Mandelstam,
Phys.\ Rep.\ {\bf 23C} (1976) 245

\bibitem{tHo76}
G. 'tHooft,
in ``High Energy Physics'', Ed. Zichichi,  Bologna 1976

\bibitem{DSS00}
H.~G.~Dosch, V.~I.~Shevchenko and Y.~A.~Simonov,
Phys.\ Rep.\ {\bf 372} (2002) 319
 arXiv:hep-ph/0007223.

\bibitem{SSP02}
A.~I.~Shoshi, F.~D.~Steffen and H.~J.~Pirner,
Nucl.\ Phys.\ {\bf A709}(2002) 131

\bibitem{DGM99}
M.~D'Elia, A.~Di Giacomo and E.~Meggiolaro,
Nucl.\ Phys.\ Proc.\ Suppl.\  {\bf 73} (1999) 515
[arXiv:hep-lat/9811017].

\bibitem{Meg99}
E.~Meggiolaro,
Phys.\ Lett.\ B {\bf 451} (1999) 414
[arXiv:hep-ph/9807567].

\bibitem{BN99}
E.~R.~Berger and O.~Nachtmann,
Eur.\ Phys.\ J.\ C {\bf 7} (1999) 459
[arXiv:hep-ph/9808320].

\bibitem{CFKS01}
A.~Capella, E.~G.~Ferreiro, A.~Kaidalov and C.~A.~Salgado,
IX Blois conference, Pruhonice 2001, Elastic and diffractive 
scattering

\bibitem{FP97}
E.~Ferreira and F.~Pereira,
Phys.\ Rev.\ D {\bf 56} (1997) 179
[arXiv:hep-ph/9705278].

\bibitem{DL98}
A.~Donnachie and P.~V.~Landshoff,
Phys.\ Lett.\ B {\bf 437} (1998) 408
[arXiv:hep-ph/9806344].

\bibitem{DDR00}
A.~Donnachie, H.~G.~Dosch and M.~Rueter,
Eur.\ Phys.\ J.\ C {\bf 13} (2000) 141
[arXiv:hep-ph/9908413].

\bibitem{DD02}
A.~Donnachie and H.~G.~Dosch,
Phys.\ Rev.\ D {\bf 65} (2002) 014019
[arXiv:hep-ph/0106169].

\bibitem{BB98}
P.~Ball and V.~M.~Braun,
Workshop on Continuous Advances in QCD, Minneapolis 1998
arXiv:hep-ph/9808229.

\bibitem{BSW87}
M.~Bauer, B.~Stech and M.~Wirbel,
Z.\ Phys.\ C {\bf 34} (1987) 103.

\bibitem{BL80}
G.~P.~Lepage and S.~J.~Brodsky,
Phys.\ Rev.\ D {\bf 22} (1980) 2157.

\bibitem{DNPW01}
H.~G.~Dosch, O.~Nachtmann, T.~Paulus and S.~Weinstock,
Eur.\ Phys.\ J.\ C {\bf 21} (2001) 339
[arXiv:hep-ph/0012367].

\bibitem{PDG02}
K.~Hagiwara {\it et al.}  [Particle Data Group Collaboration],
Phys.\ Rev.\ D {\bf 66} (2002) 010001.

\bibitem{SSDP02}
A.~I.~Shoshi, F.~D.~Steffen, H.~G.~Dosch and H.~J.~Pirner,
Phys.\ Rev.\ {\bf D66}: 094019 (2002)
arXiv:hep-ph/0207287.


\bibitem{H1-00}
C. Adloff et al., H1 Coll.
Phys.\ Lett.\ B {\bf 483} (2000) 23

\bibitem{Zeus02}
S. Chekanov et al., Zeus Coll.
Eur.\ Phys.\ J.\ C {\bf 24} (2002) 345

\bibitem{FP02} J.~R.~Forshaw and G.~Poludniowski,
arXiv:hep-ph/0107068.

\bibitem{target1}
M. Binkley et al., E-401 Coll.,
Phys.\ Rev.\ Lett. {\bf 48} (1982) 73

\bibitem{target2}
M. Arneodo et al. (NMC Coll.)
Phys.\ Lett.\ B  {\bf 332} (1994) 195

\bibitem{target3}
J. J. Aubert (EMC Coll.)
Nucl.\ Phys.\ B {\bf 213} (1983) 1

\bibitem{upsil1}
J. Breitweg et al., Zeus Coll.,
Phys.\ Lett.\ B {\bf 437} (1998) 432

\bibitem{upsil2}
L. Frankfurt, M. McDermont and M. Strikman,
JHEP\ {\bf 02} (1999) 002

\bibitem{upsil3}
A.D. Martin, M.G. Ryskin and T. Teubner,
Phys.\ Lett.\ B {\bf 454} (1999) 348

\bibitem{DD01}
A.~Donnachie and H.~G.~Dosch,
Phys.\ Lett.\ B {\bf 502} (2001) 74
[arXiv:hep-ph/0010227].







\end{thebibliography}
\end{document}